\documentclass[twocolumn,prb,aps,citeautoscript,showpacs,footinbib,amsmath,amssymb,superscriptaddress,longbibliography]{revtex4-2}
\usepackage{graphicx}
\usepackage{dcolumn}
\usepackage{xcolor}
\usepackage{bm}
\usepackage{hyperref}
\usepackage{amsmath,amssymb,amsfonts,bbold}
\usepackage[normalem]{ulem}
\usepackage{miller}
\usepackage{physics}
\usepackage{tikz}
\usepackage{array}
\graphicspath{{figures/}}

\begin{document}

\preprint{}

\title{Evidence for a conical spin spiral state in the Mn triple-layer on W(001): \\ spin-polarized scanning tunneling microscopy and first-principles calculations}

\author{Paula M. Weber}
\email[]{paula.weber@hhu.de}
\affiliation{Physikalisches Institut, Experimentelle Physik II, Universit\"at W\"urzburg, Am Hubland, 97074 W\"urzburg, Germany}
\author{Tim Drevelow}
\email[]{drevelow@physik.uni-kiel.de}
\affiliation{Institut f{\"u}r Theoretische Physik und Astrophysik, Christian-Albrechts-Universit{\"a}t zu Kiel, Germany}
\author{Jing Qi}
\email[]{jing.qi@uni-wuerzburg.de}
\affiliation{Physikalisches Institut, Experimentelle Physik II, Universit\"at W\"urzburg, Am Hubland, 97074 W\"urzburg, Germany}
\author{Matthias Bode}
\affiliation{Physikalisches Institut, Experimentelle Physik II, Universit\"at W\"urzburg, Am Hubland, 97074 W\"urzburg, Germany}
\affiliation{Wilhelm Conrad R\"ontgen-Center for Complex Material Systems (RCCM), Universit\"at W\"urzburg, Am Hubland, 97074 W\"urzburg, Germany}
\author{Stefan Heinze}
\affiliation{Institut f{\"u}r Theoretische Physik und Astrophysik, Christian-Albrechts-Universit{\"a}t zu Kiel, Germany}
\affiliation{Kiel Nano, Surface, and Interface Science (KiNSIS), Christian-Albrechts-Universit{\"a}t zu Kiel, Germany}

\date{\today}

\begin{abstract}
The spin structure of a Mn triple layer grown pseudomorphically on 
surfaces is studied using 
spin-polarized scanning tunneling microscopy (SP-STM) and density functional 
theory (DFT). 
In SP-STM images a c$(4 \times 2)$ super
structure is found. The magnetic origin of this contrast is verified by
contrast reversal and using the c$(2 \times 2)$ AFM state of the
Mn double layer as a reference.
SP-STM simulations show that this contrast
can be explained by a spin spiral propagating along the [110] direction
with an angle close to $90^\circ$
between magnetic moments of adjacent Mn rows. 
To understand the origin of this spin structure, 
DFT calculations have been performed for a 
large number of competing collinear and non-collinear magnetic states 
including the effect of spin-orbit coupling (SOC).
Surprisingly, a collinear state in which the magnetic moments of top and central
Mn layer are aligned antiparallel and those of the bottom Mn layer are aligned parallel 
to the central layer is the energetically lowest state. We show that in this so-called
``up-down-down" ($\uparrow \downarrow \downarrow$) state the magnetic moments in the Mn bottom layer 
are only induced by those of the central Mn layer.
Flat spin spirals propagating either in one, two, or all Mn layers are shown
to be energetically unfavorable to the collinear $\uparrow \downarrow \downarrow$ state
even upon including the Dzyaloshinskii-Moriya interaction (DMI). 
However, conical spin spirals with a small opening angle of about $10^\circ$
are only slightly energetically unfavorable within DFT and could explain the experimental observations.
Surprisingly, the DFT energy dispersion of conical spin spirals 
including SOC cannot be explained if only the DMI is taken into account. Therefore,
higher-order interactions such as chiral biquadratic terms need to be considered
which could explain the stabilization of a conical spin spiral state.
\end{abstract}


\maketitle

\section{Introduction}
In ultrathin transition-metal films on surfaces, a great variety of
intriguing magnetic states has been observed with atomic resolution
using spin-polarized 
scanning tunneling microscopy (SP-STM), such as two-dimensional antiferromagnets 
\cite{Bluegel1988,Heinze2000,Kubetzka2005,meyer2020}, N\'eel states \cite{Gao2008,Wasniowska2010},
flat and conical spin spiral states \cite{Bode2007a,ferriani2008,Yoshida2012}, 
multiple-Q states \cite{Kroenlein2018,Spethmann2020,Gutzeit2022}, 
chiral domain walls \cite{Pietzsch2001,Meckler2009,Perini2019}, or magnetic skyrmions \cite{Heinze2011,Romming2013,Herve2018,meyer2019}. 
The study of such structurally well-defined model systems allows to understand 
complex spin structures based on
the underlying magnetic interactions. In this way, the interfacial Dzyaloshinskii-Moriya
interaction (DMI) has been discovered \cite{Bode2007a,ferriani2008} and higher-order exchange
interactions have been revealed which can lead to three-dimensional spin structures \cite{Heinze2011,Yoshida2012,Kroenlein2018,Spethmann2020}. 

So far, most of these studies have focused on systems consisting of one or two atomic
layers of a magnetic material, such as Mn, Fe, or Co on a metallic surface. For spintronic devices,
on the other hand, thicker film structures with at least a few atomic layers of 
a magnetic $3d$ transition-metal are required which are interfaced with non-magnetic metallic layers. 
Therefore, it is interesting to extend studies at surfaces to systems with multiple magnetic layers.
However, due to the lattice mismatch between the substrate and the magnetic film, such
systems often exhibit complex structural relaxations and superstructures 
\cite{Finco2016,Finco2017,Hauptmann2018,Dupe2018}
that are hard to take into account in density functional
theory (DFT) calculations. 
As a result, in such film systems an understanding of the spin structure 
and its origin from first-principles electronic structure theory is often limited. 

In particular, for ultrathin Mn films on the (001) surface of body-centered cubic (bcc) tungsten (W) 
a large body of theoretical \cite{ferriani2005, dennler2005, dennler2005_diffusion, ondracek2007} 
and experimental studies \cite{tian2001, ferriani2008, meyer2020} is available.   
Even for the monolayer (ML) and the double-layer (DL) these
investigations revealed complex spin structures. 
In 2005, Dennler and Hafner proposed pseudomorphic growth of the Mn ML and DL on 
W(001) based on DFT calculations
and predicted a ferromagnetic ground state due to hybridization with the substrate \cite{dennler2005}. The ferromagnetic ground state of the Mn ML was found
independently by Ferriani {\it et al.} at the same time via DFT 
calculations \cite{ferriani2005}. 
Indeed, low-energy electron diffraction (LEED) and Auger electron spectroscopy (AES) experiments 
confirmed the pseudomorphic growth of Mn on W(001) up to a film thickness of $\approx 16$\,\AA,  
resulting in the body-centered tetragonal $\delta$-phase which adopts the lateral lattice constant 
of W(001), $a_{\rm W} = 316.5$\,pm \cite{tian2001}. 

Intriguing magnetic properties, dependent on film thickness, have been experimentally 
reported for Mn/W(001). Ferriani and co-workers revealed that the Mn ML on W(001) 
exhibits a spin spiral state using SP-STM and explained this discrepancy to the
predicted ferromagnetic state based on DFT as a result of the DMI \cite{ferriani2008}
which occurs due to spin-orbit coupling not taken into account
in the earlier DFT calculations.
For the Mn DL on W(001), a collinear
antiferromagnetic (AFM) order with a magnetic c($2 \times 2$) unit cell 
and out-of-plane easy magnetization axis was observed \cite{meyer2020}, in contradiction
to the predictions of Dennler {\it et al.} \cite{dennler2005}. 
Interestingly, DFT calculations performed in latter study revealed 
that the interfacial Mn layer is magnetically dead, i.e., it carries a vanishing magnetic moment, due to the strong
hybridization with the W surface \cite{meyer2020}. 
Theoretically, Dennler {\em et al.} predicted pseudomorphic growth and the transition from a ferromagnetic to an antiferromagnetic interlayer 
exchange coupling when going from the Mn DL to the Mn triple-layer 
\cite{dennler2005}. However, experimental results have not been 
available so far.

Here, we explore the structural and magnetic properties of a Mn triple-layer (3L) on W(001) by combining
SP-STM experiments with SP-STM simulations and DFT calculations.
Experimentally we find that 3L Mn indeed grows pseudomorphic on W(001).  
SP-STM reveals a
$(2 \sqrt{2} \times \sqrt{2})$ magnetic unit cell which can be consistently explained by a flat or a conical $90^\circ$ spin spiral propagating
along the [110] direction as shown by simulations of SP-STM images.
Attempts to verify this spin structure as the magnetic ground state by DFT turned out to be highly intricate. 
Various spin structures, such as co-planar layered magnetic, antiferromagnetic, flat cycloidal or conical spin spirals, and a superposition of two 90° spin spirals with opposite rotational sense (the so-called $uudd$ state) are compared. 
In contrast to the work of Dennler {\em et al.} \cite{dennler2005}, we find that 
the energetically lowest collinear magnetic state exhibits an antiparallel alignment of the magnetic moments in the two upper Mn layers while the 
magnetic moments of the Mn interface layer are 
parallel to those of the central Mn layer.

Based on spin spiral calculations we show that the exchange interactions in 
the Mn triple layer are frustrated due to competing antiferromagnetic exchange couplings 
between and within the upper two Mn layers, an effect which is influenced by the Mn interlayer distances. 
In particular, we find that interlayer exchange prefers a 
collinear spin alignment while the intralayer exchange favors a spin spiral state.
The DMI naturally promotes cycloidal spin spiral states. However, its energy 
contribution turns out to be rather small which we attribute to the small induced 
magnetic moments of the Mn layer at the interface with the W substrate. The
magnetocrystalline anisotropy as well as the magnetic dipole interaction are 
only about 0.05~meV/Mn atom and favor an in-plane magnetization. $90^\circ$ conical
spin spiral states with a small opening angle -- which can explain the SP-STM 
experiments -- are still slightly higher in total energy than the collinear 
$\uparrow \downarrow \downarrow$ state. Surprisingly, we find that these DFT 
calculations can only be explained if we take higher-order interactions due to
spin-orbit coupling into account such as the chiral biquadratic pair interaction.

\section{Experimental methods}
\label{sec:exp_methods}
All experiments were performed in a two-chamber ultra-high vacuum (UHV) system 
with a base pressure $p \leqslant 1 \times 10^{-10}$\,mbar. 
Clean W(001) was prepared in the preparation chamber by numerous cycles 
consisting of $5$\,min annealing at $T_{\rm ann} = (1580 \pm 50)$\,K 
in an oxygen atmosphere, followed by an about $12$\,s 
long high-temperature flash at $T_{\rm fl} = (2400 \pm 100)$\,K.  
To remove potential carbon from the surface and to avoid unwanted oxidation of W at the same time, 
we successively reduced the oxygen pressure from $p_{\rm O_{2}} \simeq 5 \times 10^{-8}$\,mbar 
in the initial cycle to $p_{\rm O_{2}} \simeq 1 \times 10^{-9}$\,mbar in the final cycle \cite{bode2007}.  

After the final cycle, the oxygen dosing valve was closed, and the W(001) crystal was flashed again for 15\,s. 
Once the pressure dropped to $p < 3 \times 10^{-10}$\,mbar, the Mn-loaded crucible 
of a commercial high-temperature effusion cell evaporator was preheated to a nominal temperature of $966.5$\,K 
for about 3\,min to stabilize the evaporator and the pressure. 
Mn deposition onto the W(001) substrate was started at a sample temperature $T_{\rm s} \approx 333$\,K. 
During evaporation, the pressure indicated by the gauge was $p < 1 \times 10^{-9}$\,mbar. 
After Mn deposition, the films were annealed at $493 \pm 20$\,K for $14$\,min $\pm \, \, 4$\,min.
All Mn coverages mentioned below are given in pseudomorphic atomic layers (p-AL) on W(001).

Immediately after preparation, the crystal was transferred into a home-built low-temperature STM 
housed in a UHV-compatible liquid He cryostat ($T_{\rm STM} = 4.5$\,K).  
We used electrochemically etched polycrystalline W tips.  
For spin-resolved STM measurements, these W tips were magnetized \textit{in situ} 
by gentle poking the tip apex into a Mn film ($\approx 500$\,pm) and pulsing ($\approx 10$\,V), 
similar to description in Ref.\,\onlinecite{loth2010}. 
All STM images were processed using WSxM \cite{Horcas2007}.

\section{Computational details}
\label{sec:comp_details}
We studied the structural, electronic, and magnetic properties 
of the Mn triple layer on W(001) using DFT.
For structural relaxations of collinear magnetic states and 
to calculate the energy dispersion of spin spiral states 
we used the full-potential linearized augmented plane-wave (FLAPW) method
as implemented in the {\tt FLEUR} code \cite{FLEUR,Kurz2004,Zimmermann2014}.
The relaxation of non-collinear magnetic
states and all calculations in the $c(4\times2)$ supercell were carried
out with the projected augmented wave (PAW) method as
implemented in the {\tt VASP} code 
\cite{VASP,Kresse96,Kresse99}.

Structural relaxations with the {\tt FLEUR} code were performed 
for collinear magnetic states
in the generalized gradient approximation (GGA) using the
exchange-correlation potential by Perdew and Wang \cite{pw91}.
We used the theoretical GGA lattice constant of W which amounts to
$a=3.17$~{\AA} and differs from the experimental lattice constant by only 0.5\%.
66 $k$-points were used in the irreducible part of the
two-dimensional Brillouin zone. In these calculations for collinear magnetic
states a symmetric film with a total of 9 W layers and three Mn layers on either side was applied. The top three Mn and top two W layers were relaxed in the 
direction perpendicular to the film
until the forces on each atom were below 0.001~htr/a.u. The muffin-tin spheres of the Mn and W atoms had radii of 2.3~a.u. and 2.5~a.u. respectively. The plane wave cutoff parameter was set to 
$k_{\rm max}=4.0$~a.u.$^{-1}$ and the 5$p$ semicore states of 
W were described by local $p$ orbitals.
The relaxation of the checkerboard antiferromagnetic state was done in a $\text{c}(2\times2)$ supercell on a $17\times17$ k-point grid with the $\uparrow\downarrow\uparrow$-state as a reference.

For spin spiral calculations in {\tt FLEUR}, we used an asymmetric film consisting
of 9 W layers and a Mn triple
layer on only one side of the film. The relaxed interlayer distances from the calculations
for collinear magnetic states were used. 
The number of $k$-points was increased to 2304 in the entire 
Brillouin zone and the exchange-correlation potential was treated
in local density approximation (LDA) using the parametrization of Vosko, Wilk, and Nusair \cite{vwn}.
The plane wave cutoff parameter
$k_{\rm max}=4.0$~a.u.$^{-1}$ and other settings remained unchanged from structural relaxation.
The contribution of the DMI to the energy dispersion of spin spirals was 
obtained treating SOC in first order perturbation theory \cite{Heide09}
since a self-consistent treatment of SOC is incompatible 
with the generalized Bloch theorem used for spin spirals.
The calculation of the magnetocrystalline anisotropy energy was carried out 
in the same atomic setup including spin-orbit-coupling (SOC) self-consistently \cite{Li1990}
and 18225 k-points in the total Brillouin zone with a plane wave cutoff parameter
$k_{\rm max}=4.0$~a.u.$^{-1}$.

Structural relaxations in {\tt VASP} were performed in a 
non-collinear setup to allow the relaxation of spin spiral states. 
A GGA exchange correlation functional was used \cite{pbe}. 
Calculations were carried out in the chemical unit cell using spin spiral boundary conditions, 
on a $(22\times22)$ $k$-point grid. An energy cutoff parameter of 300 eV was chosen for the plane-wave basis set. 
Atoms were arranged in the same symmetric slab with 9 W layers compared with {\tt FLEUR} calculations, 
where the top 5 layers are free to relax into the $z$-direction, i.e.~perpendicular to the surface. 
Collinear states were also relaxed to check consistency with the results obtained from {\tt Fleur}. 
For total energy calculation in the $c(4\times2)$ supercell the LDA exchange-correlation potential 
by Vosko, Wilk, and Nusair \cite{vwn} was used together with an energy cutoff of 300 eV 
and a $(12\times24)$ grid of $k$-points. 
The substrate was modeled asymmetrically by 9 W layers with 3 Mn layer on one side. 
For calculations with spin-orbit coupling (SOC), the energy cutoff was increased to 390 eV
and the size of the k-point grid was increased to $(18\times36)$.

\section{Results}
	\subsection{Experimental results}

\begin{figure*}[t]
\includegraphics[scale=1]{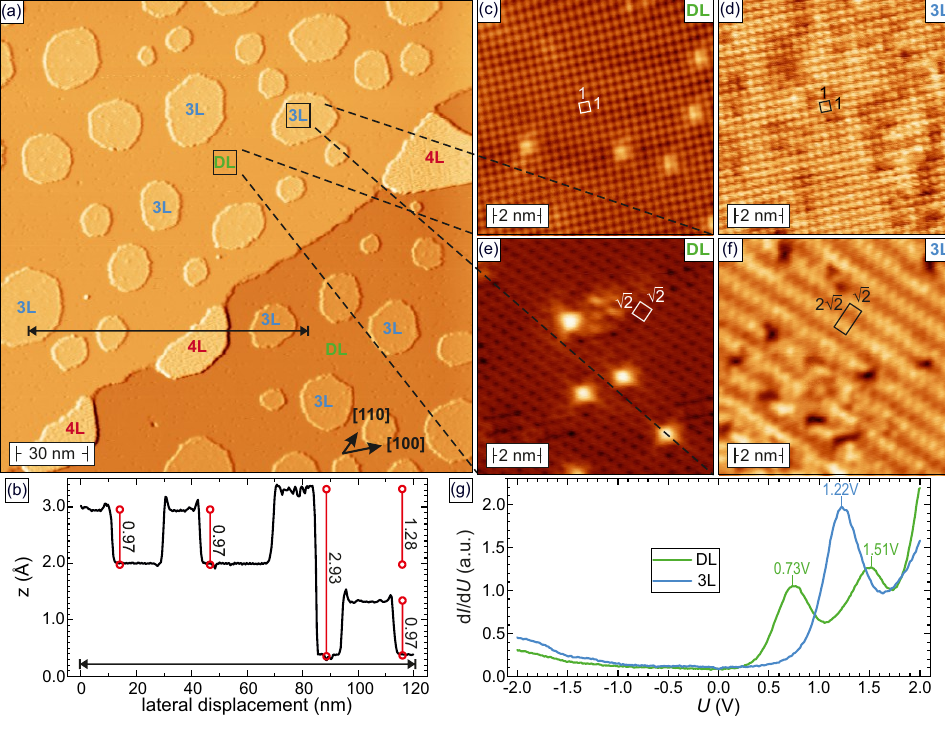}
\caption{\label{fig:overview}(a) Overview STM topographic scan of a Mn film on W(001) 
		with an average coverage $\theta_{\rm Mn}^{\rm av} = (2.3 \pm 0.2)$\,p-AL measured with a magnetic Mn/W tip.
		(b) Line profile measured between the two black arrows in (a). 
		Spin-averaged atomic-resolution STM data taken on (c) a DL Mn terrace and (d) a 3L Mn island. 
		In both cases a square-shaped ($1 \times 1$) unit cell is observed, indicating pseudomorphic growth.  
		(e,f) Atomically spin-resolved SP-STM scans of DL and 3L Mn, respectively. 
		A ($ \surd 2 \times \surd 2$) magnetic unit cell is reproduced for DL in (e) (compare with Ref.~\cite{meyer2020}), 
		whereas a $2\surd 2 \times \surd 2$ structure is observed for 3L Mn in (f). 
		(g) Tunneling spectra measured on DL Mn terraces (green) and 3L Mn islands (blue). 
		Scan parameters: (a) \mbox{$U = 1\,\mathrm{V}$,} \mbox{$I = 300\,\mathrm{pA}$};  
		(c,f) \mbox{$U = 10\,\mathrm{mV}$,} \mbox{$I = 4\,\mathrm{nA}$};  
		(d,e) \mbox{$U = 10\,\mathrm{mV}$,} \mbox{$I = 5\,\mathrm{nA}$};
		Stabilization parameters: (g) \mbox{$U = 2\,\mathrm{V}$,} \mbox{$I = 300\,\mathrm{pA}$}.}
\end{figure*}

\begin{figure*}[thb]
\includegraphics[scale=1]{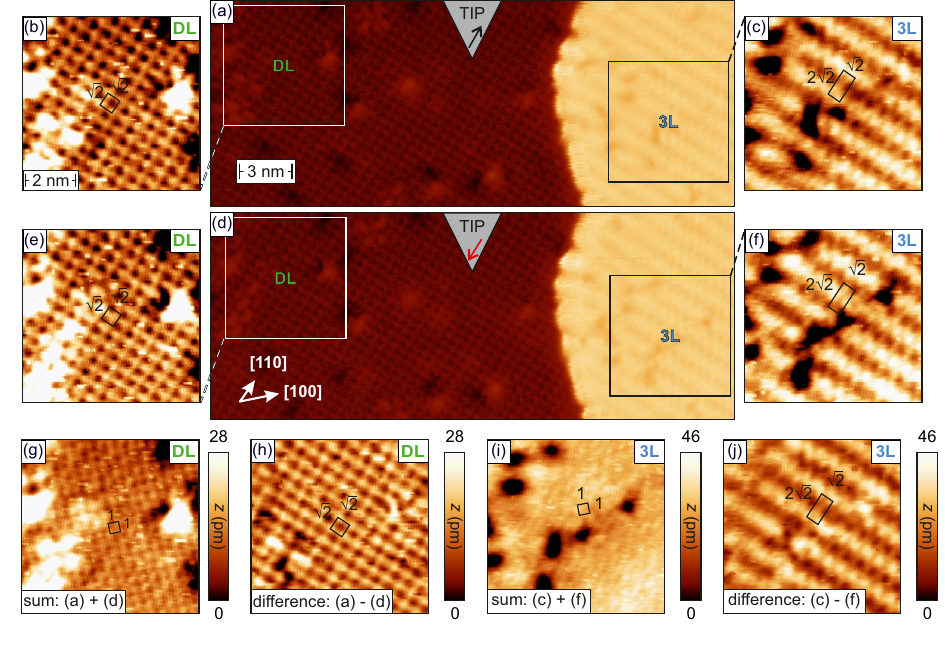}
\caption{\label{fig:contrastreversal} (a) SP-STM image of a surface area covered 
        by a double-layer (DL) and triple-layer (3L) Mn/W(001).  
		(b) Zoomed image of the DL showing the well-known AFM $\surd 2 \times \surd 2$ magnetic unit cell \cite{meyer2020}. 
		(c) Zoomed image of 3L Mn/W(110) with the $2\surd 2 \times \surd 2$ magnetic unit cell. 
		(d) SP-STM scan of the same area as shown in (a) but now measured with a reversed out-of-plane magnetic Mn/W tip.  
		(e,f) Magnified views showing the magnetic structure of the Mn DL and 3L, respectively.
		(g) Sum and (h) difference of the images (b) and (e) taken on the Mn DL before and after tip magnetization reversal, respectively.  
		Note that while (g) shows the structural unit cell and emphasizes the topographic signature of defects, 
		(h) is dominated by the magnetic unit cell and defects are almost invisible.  
		(i) Sum and (j) difference of the images (c) and (f) taken on 3L Mn before and after tip magnetization reversal, respectively.   		
		Scan parameters: (a,d) \mbox{$U = 10\,\mathrm{mV}$,} \mbox{$I = 4\,\mathrm{nA}$};} 
\end{figure*}

The experimental results of our (SP)-STM experiments are summarized in Fig.~\ref{fig:overview}.
All data were taken on a sample which was coated 
with an average Mn coverage $\theta_{\rm Mn}^{\rm av} = (2.3 \pm 0.2)$\,p-AL.
The 200\,nm\,$\times\,200$\,nm overview scan in Fig.~\ref{fig:overview}(a) reveals smooth terraces 
which are covered by a Mn double-layer (DL) with numerous roughly circularly shaped 
Mn triple-layer (3L) islands on top.  
The typical island diameter amounts from about $7$\,nm up to $30$\,nm. 
The line section plotted in Fig.~\ref{fig:overview}(b) measured between the two black arrows in Fig.~\ref{fig:overview}(a) 
shows an apparent height of $h_{\rm app}^{\rm 3L} \approx (100 \pm 20)$\,pm at the tunneling parameters chosen here, 
i.e., a sample bias voltage \mbox{$U = 1\,\mathrm{V}$} and a tunneling current \mbox{$I = 300\,\mathrm{pA}$}. 
Moreover, semi-elliptical islands of the fourth Mn layer (4L) with diameters between $15$\,nm (perpendicular to the step edge) 
and $110$\,nm (along the step edge) and a height $h_{\rm app}^{\rm 4L} \approx  (290 \pm 20)$\,pm can be found at step edges.

Atomic-resolution measurements performed with a non-magnetic W tip on the flat terrace (covered by a Mn DL) 
and 3L
Mn islands are presented in Fig.~\ref{fig:overview}(c) and (d), respectively. 
As indicated by white and black boxes, both data sets exhibit a square-shaped $(1 \times 1)$ unit cell 
with a lattice constant consistent with the underlying W(001) substrate, $a_{\rm W} = 316.5$\,pm.
Magnetically sensitive SP-STM data acquired on surface areas covered by Mn DL or 3L
[indicated in Fig.~\ref{fig:overview}(a)] are shown in Fig.~\ref{fig:overview}(e) and (f), respectively.  
For the former, we recognize a $\surd 2 \times \surd 2$ unit cell 
which is rotated by $45^{\circ}$ with respect to the $[100]$ direction of the W(001) substrate. 
The contrast is dominated by holes (depressions) surrounded by a grid of linear elevations.
As discussed in detail in Ref.~\onlinecite{meyer2020}, this mesh-like appearance in STM images 
is characteristic for the out-of-plane antiferromagnetic Mn DL on W(001).  

Fig.~\ref{fig:overview}(f) presents typical atomic-resolution SP-STM data taken on a 3L Mn island. 
Zigzag--shaped stripes with an inter-stripe separation of $(1.03 \pm 0.14)$\,nm 
and a periodicity of $(0.49 \pm 0.05)$\,nm along the stripes are observable, 
corresponding to a $2\surd 2 \times \surd 2$ magnetic unit cell indicated by a black rectangle. 
As shown in Fig.~\ref{fig:overview}(g), by recording local tunneling spectra, 
we detect a very different local electronic structure for DL and 3L Mn.  
Whereas the DL exhibits peaks at $U = +0.73$\,V and $U = +1.51$\,V (green curve),
the tunneling spectrum of 3L Mn yields only one characteristic peak at $U = +1.22$\,V.

To unambiguously confirm the magnetic origin of the zigzag--shaped 
contrast observed in Fig.~\ref{fig:overview}(f) 
we performed experiments in which the same locations covered by DL and 3L Mn on W(001) 
were imaged by SP-STM before and after reversing the tip magnetization, see Fig.~\ref{fig:contrastreversal}.  
A high-resolution SP-STM scan of a surface area 
covered by a Mn DL in the darker left part and by a 3L Mn film on the brighter right part is shown in Fig.~\ref{fig:contrastreversal}(a). 
On both surface areas a significant magnetic contrast is obtained, 
as highlighted in the zoomed-in images presented in Fig.~\ref{fig:contrastreversal}(b) and (c), respectively.  
Since it is known from DFT calculations that the spin structure of the Mn DL on W(001) is out-of-plane antiferromagnetic \cite{meyer2020},
we can safely conclude that the tip magnetization must exhibit a significant out-of-plane component.  
Yet, the absolute magnetization direction of the tip is unknown and the scheme 
in the upper part of Fig.~\ref{fig:contrastreversal}(a) serves illustrative purposes only.  
We also note that we cannot exclude that this out-of-plane magnetization coexists with an in-plane component.

To reverse the magnetization of the out-of-plane component of the tip magnetization, 
the Mn-W tip was carefully approached by a distance $\Delta z$ towards the Mn film. 
To exclude any effect on the previously imaged surface areas, this approach was performed about $23$\,nm below the region shown in Fig.~\ref{fig:contrastreversal}(a). 
At $\Delta z \approx 250$\,pm, we suspect that tip and sample orbitals overlap at such close tip--sample distance, 
resulting in a significant exchange interaction
\cite{tao2009,hsu2010,Schmidt2011,Schmidt2012,Hauptmann2020}
which magnetically reverses the Mn cluster at the tip apex.

Upon this procedure we moved back to the position of Fig.~\ref{fig:contrastreversal}(a) 
and scanned the same surface area with a reversed tip magnetization, see Fig.~\ref{fig:contrastreversal}(d). 
Zoomed-in scans of the Mn DL and 3L are shown in Fig.~\ref{fig:contrastreversal}(e) and (f), respectively. 
Again the characteristic $\surd 2 \times \surd 2$ and $2\surd 2 \times \surd 2$ magnetic unit cells are observable.
By using defect sites as markers to precisely aligning the images shown in panels (b) and (e) for the DL, 
a half-period shift of the magnetic contrast becomes evident.  
This first impression is corroborated by calculating the sum (g) and the difference (h) of the DL images displayed in (b) and (e).
Whereas the sum in Fig.~\ref{fig:contrastreversal}(g) exhibits the square-shaped $(1 \times 1)$ structural unit cell, 
the difference (h) demonstrates the familiar AFM $\surd 2 \times \surd 2$ magnetic unit cell, in agreement with Ref.~\onlinecite{meyer2020},  
thereby unambiguously confirming the reversal of the out-of-plane component of the tip magnetization.  

Application of the same procedure to the SP-STM data presented in Fig.~\ref{fig:contrastreversal}(c) and (f), which were obtained on the 3L region, 
results in the images presented in Fig.~\ref{fig:contrastreversal}(i) for the sum and (j) for the difference.  
While the sum (i) yields a blurred signal only, possibly due to the very low structural atomic corrugation amplitude of 3L Mn on W(001),
the difference image presented in panel (j) clearly reveals the magnetic $2\surd 2 \times \surd 2$ unit cell, which can also be referred to as the magnetic c$(4 \times 2)$ unit cell.  
Please note that the defects are hardly visible in (j) but very pronounced in (i), 
indicating that the difference image tends to cancel topographic and highlight magnetic contrasts.

\subsection{SP-STM simulations}

\begin{figure}
    \centering
    \includegraphics[width=0.485\textwidth]{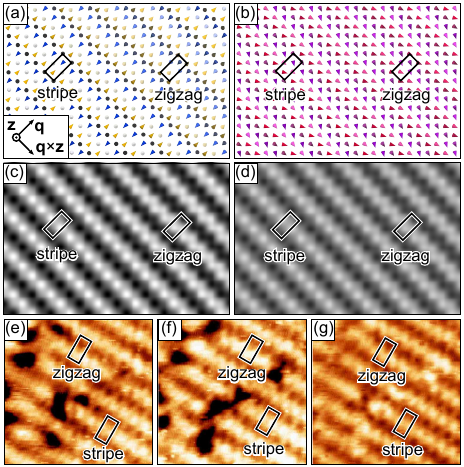}
    \caption{(a-d): Simulation of SP-STM images of a flat spin spiral (left
    column) 
        and a conical spin spiral (right column), propagating into the direction \textbf{q} indicated and exhibiting an angle of 87.5° between the 
        magnetic moments of Mn atoms in adjacent rows of the top layer. 
        (a,b) Atomic magnetic moments in the top Mn layer. 
        (c,d) Simulated magnetic contrast of SP-STM images according to the 
        Tersoff-Hamann model for a magnetic tip with a 
        magnetization direction perpendicular to the surface, i.e.~in
        the $z$-direction with normalized contrast. 
        In both simulated images a zigzag domain and a stripe domain can be found, 
        which are also observed in the SP-STM images displayed in (e) and (f) as well as in their difference in (g). Note that panels (e-g) are the same as 
        panels (c), (f), and (j) from Fig.~\ref{fig:contrastreversal}.}
    \label{fig:STMsimulation}
\end{figure}

The $c(4\times2)$ super cell suggested by the experimental SP-STM images as the magnetic unit cell of the Mn triple layer can be
explained by
a 90° spin spiral state propagating along the [110] direction
of the surface. 
In such a state the magnetic moments of Mn atoms in adjacent
rows perpendicular to the propagation direction are rotated by
$90^\circ$ and the spiral completes a rotation after four lattice 
sites which exactly matches the $c(4\times2)$ super cell.
To check whether the
experimentally observed contrast can be reproduced by assuming such a spin spiral state, 
we have simulated SP-STM images (Fig.~\ref{fig:STMsimulation}) based on  
the spin-polarized generalization \cite{Wortmann2001} of the
Tersoff-Hamann model \cite{tersoffhamann1985} using the 
the model described in Ref.~\cite{heinze2006}. Note, that
in these SP-STM simulations only the magnetic moment directions of 
the Mn atoms in the surface layer enter, whereas the other two Mn layers
have no impact on the simulated images.

For the SP-STM simulations we actually chose an angle of 87.5° between the magnetic moments of Mn atoms in adjacent rows of the top layer (Fig.~\ref{fig:STMsimulation}(a))
instead of 90° as this leads to an even better matching of the experimental images. In particular, we observe
the formation of different rotational domains (contrasts)
on a larger scale while
leaving the spin structure locally close to the 90° spiral.
The two domains are denoted as stripe and zigzag domains (Fig.~\ref{fig:STMsimulation}(a,c)). In the stripe domain the
Mn magnetic moments and the magnetization direction of the STM tip enclose angles of 0°, 90° or 180°, while in the other domain a 
zigzag contrast is found due to enclosing angles of about 45° 
and 135° (Fig.~\ref{fig:STMsimulation}(e)). These two types of
contrast also appear in the experimental images 
(Fig.~\ref{fig:STMsimulation}(e-g)).

However, a conical spin spiral with a finite opening angle
and an angle of 87.5° between the flat spin spiral component
of the magnetic moments of Mn atoms in adjacent rows 
Fig.~\ref{fig:STMsimulation}(b,d) can
explain the experimental observations as well.
SP-STM simulations reveal that the magnetic contrast only
changes in magnitude when going from a flat towards a conical spiral 
(Fig.~\ref{fig:STMsimulation}(c) vs.~(d)). 
The magnetic contrast of conical spirals cannot be distinguished from a flat spin spiral 
as long as the magnetization direction of the STM tip and the direction of the in-plane
component
are perpendicular to each other. This is consistent with the experimental 
situation since the easy magnetization direction of the
$\uparrow\downarrow\downarrow$ state obtained in DFT 
is in the plane of the film (see section \ref{sec:DFT}) and
the STM tips had an out-of-plane magnetization direction as deduced from
the contrast on the Mn double layer.

\subsection{First-principles calculations}
\label{sec:DFT}

\begin{figure*}[t]
\includegraphics[width=\textwidth]{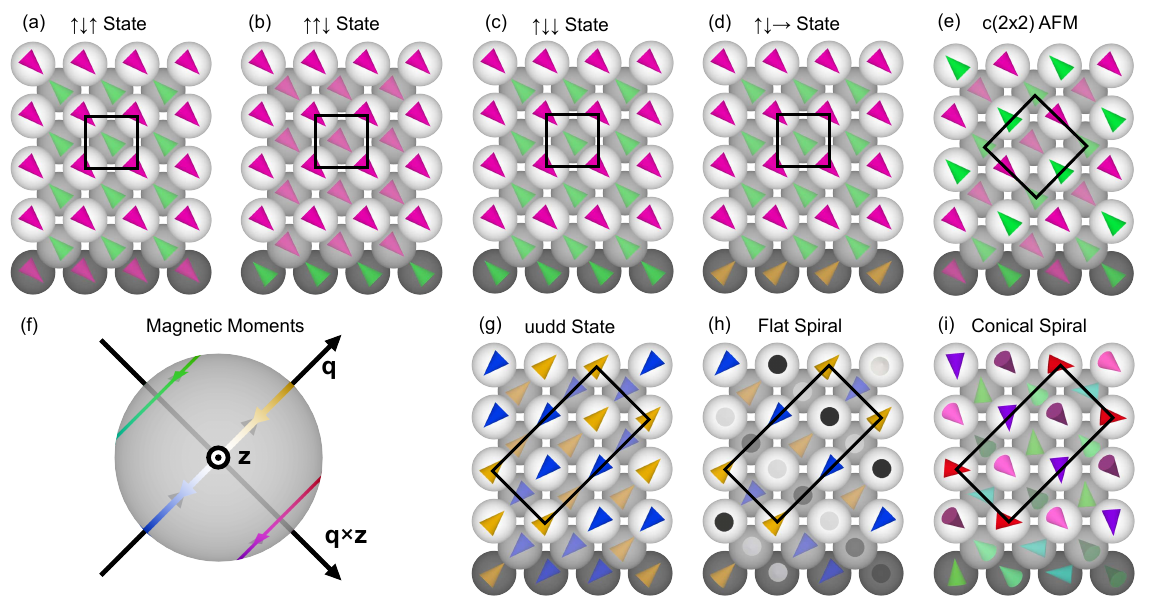}
    \caption{Sketches of selected magnetic configurations investigated in the DFT study. 
         The magnetic moments are indicated for Mn atoms in the top layer 
         (white spheres), 
         in the central layer (light gray spheres), and in the bottom 
         layer (dark gray spheres). 
         (a-e) Co-planar magnetic states with
         parallel magnetic moments
         in each layer. Note that spin states (a-d)
         do not lead to the observation of a magnetic superstructure in 
         SP-STM images. 
         The spin structures differ in the magnetic alignment of the bottom Mn layer.
         Spin spirals lie a the intersection with a cone. The arrows show the sense of rotation of the moments when moving along \textbf{q}, 
         as it is preferred by the Dzyaloshinskii-Moriya interaction.
         (f) Stereographic projection of the magnetic moments from the flat and conical spin spiral onto a sphere. Arrows indicate the rotational sense of the spiral as it propagates along the direction of $\textbf{q}$.
         (g) $uudd$-state that leads to a c(4$\times$2) magnetic superstructure and is created by a superposition of two $90^\circ$ spin spirals with opposite rotational sense.
         (h) Flat cyclodial spin spiral with an angle between adjacent moments within a Mn layer close to $90^\circ$.
         (i) Conical spin spiral with an opening angle of 40° which is a superposition of the states in (c) and (h).}
    \label{fig:magneticStates}
\end{figure*}

\begin{table*}[t]
    \centering
    \begin{tabular}{ccccccccccc}
         \hline \hline
         state & method & $d_{\text{T}/\text{M}}$ & $d_{\text{M}/\text{B}}$ & $d_{\text{B}/\text{1}}$ & $d_{\text{1}/\text{2}}$ & $d_{\text{2}/\text{3}}$ 
         & $\Delta E$ & $m_\text{T}$ & $m_\text{M}$ & $m_\text{B}$ \\\hline 
         $\uparrow\downarrow\uparrow$ 
         & {\tt FLEUR}
         & $-$19.5 &  $-$5.5 & $-$10.9 &  1.0 & 2.3 
         & 0 & 3.71 & $-$3.04 & 2.36 \\
         $\uparrow\downarrow\downarrow$ 
         & {\tt FLEUR}
         & $-$32.8 &  $-$6.5 & $-$16.1 &  4.3 & 1.0 
         & $-$122 & 3.45 & $-$2.92 & $-$1.27\\
         $\uparrow\uparrow\downarrow$ 
         & {\tt FLEUR}
         &  $-$7.3 & $-$23.9 &  $-$2.5 & $-$1.6 & 3.2 
         & 35 & 3.75 & 2.37 & $-$2.39\\
         AFM 
         & {\tt FLEUR}
         &  $-$17.7 & $-$15.1 &  $-$13.0 & 2.9 & $-$1.64 
         & $-$36 & $\pm$3.70 & $\pm$2.87 & $\mp$0.90\\\hline
         $\uparrow\downarrow\uparrow$   
         & {\tt VASP}
         & $-$24.4 &  $-$4.0 & $-$15.8 &  0.5 & $-$0.6
         & 0 & 3.67 & $-$3.13 & 1.90 \\
         $\uparrow\downarrow\downarrow$ 
         & {\tt VASP}
         & $-$31.9 &  $-$7.8 & $-$16.4 &  2.1 & $-$0.5
         & $-$145 &  3.28 & $-$2.79 & $-$1.40\\
         $\uparrow\uparrow\downarrow$ 
         & {\tt VASP}
         & $-$8.3 &  $-$24.7 & $-$3.8 &  $-$3.4 & 0.9
         & 8 &  3.72 & 2.36 & $-$2.39\\
         $\overline{\Gamma'\text{M}}$ 
         & {\tt VASP}
         & $-$27.2 &  $-$9.8 & $-$15.7 &  1.6 & $-$0.5
         & $-$90 &  $-$3.47 & $-$2.90 & $-$1.35\\\hline
         $\uparrow\downarrow\uparrow$ 
         & Ref.~\cite{dennler2005} 
         & $-20.0$ &  $-5.4$ & $-12.6$ & $-0.6$ & $-0.5$
         & 0 & 3.78 & $-$3.14 & 2.39 \\
         $\uparrow\uparrow\uparrow$  
         & Ref.~\cite{dennler2005} 
         & $-13.1$ & $-21.7$ & $-14.5$ &  0.7 & $-0.8$
         & 97 & 3.71 & 1.34 & 1.43 \\
         \hline \hline
    \end{tabular}
    \caption{Results of structural relaxations 
    of the Mn triple layer on W(001)
    for selected magnetic states  (cf.~Fig.~\ref{fig:magneticStates}) 
     obtained via the {\tt FLEUR} and {\tt VASP} code.    
     In the table the c$(2 \times 2)$ AFM state has been
     abbreviated by AFM
     and the $90^\circ$ spin spiral along 
     $\overline{\Gamma'\text{M}}$ by $\overline{\Gamma'\text{M}}$.
     The relative changes $d$ in \% of the interlayer distances      
     with respect to the W bulk interlayer distance of $1.59$~{\AA}
     are given between the top (T), middle (M) and bottom (B) Mn layer as well as between
     the three top W layers. The difference in total energy $\Delta E$ relative to the 
     $\uparrow\downarrow\uparrow$-state is given in meV/Mn atom. The magnetic moments of 
     the Mn atoms are given in $\mu_\text{B}$.
     For comparison the results of 
     Dennler \textit{et al.} \cite{dennler2005} for the states considered in that work
     are given in the last two lines.}
    \label{tab:relaxation}
\end{table*}

{\bf Structural relaxations and collinear states.}
First we performed DFT calculations for three collinear magnetic states to determine the exchange coupling %
between the upper and lower two Mn 
layers (Fig.~\ref{fig:magneticStates}(a-c)). The investigated states were  
the layerwise antiferromagnetic 
($\uparrow\downarrow\uparrow$) state (Fig.~\ref{fig:magneticStates}(a)), 
and two states in which the magnetic moments of two adjacent Mn layers are 
parallel and antiparallel with respect to the other Mn layer, i.e.~the
$\uparrow\uparrow\downarrow$ 
(Fig.~\ref{fig:magneticStates}(b))
and the $\uparrow\downarrow\downarrow$
state (Fig.~\ref{fig:magneticStates}(c)), where each arrow denotes the magnetization of one Mn layer from top to bottom. Note, that
we could not converge the ferromagnetic ($\uparrow\uparrow\uparrow$) state 
in our DFT calculations
as the magnetic moments of the Mn layers flipped
such that we arrived at the $\uparrow \downarrow \downarrow$ state.

We carried out structural relaxations for each of the collinear magnetic
states using both the {\tt FLEUR} and the {\tt VASP} code which
give similar results (Tab.~\ref{tab:relaxation}). Our results show
that there is a large inward relaxation of the Mn layers which depends
sensitively on the considered magnetic state.
For the layerwise antiferromagnetic ($\uparrow\downarrow\uparrow$) state, the relaxations and
magnetic moments are in good agreement with those of a previous DFT study
on the Mn triple layer on W(001) by Dennler \textit{et al.}~\cite{dennler2005},
as shown in Tab.~\ref{tab:relaxation}. 
Note, that only the ferromagnetic and the $\uparrow\downarrow\uparrow$ 
state have been considered in that work \cite{dennler2005}.

Surprisingly, the $\uparrow\downarrow\downarrow$ state (Fig.~\ref{fig:magneticStates}(c)) is energetically much more favorable than the $\uparrow\downarrow\uparrow$ state in both the {\tt FLEUR}
and the {\tt VASP} calculation (Tab.~\ref{tab:relaxation}). 
This indicates antiferromagnetic coupling between magnetic moments of the 
two upper Mn layers and ferromagnetic coupling between the Mn moments of the central and bottom layer. The magnetic moment of the interfacial (bottom) Mn layer is
much smaller than that of the other two Mn layers. The magnetic moment of the
bottom Mn layer vanishes if the moment direction 
is constrained within the {\tt FLEUR} calculations
to be perpendicular to the magnetic moments of the two upper layers (Fig.~\ref{fig:magneticStates}(d)). This indicates that the Mn atoms at the W interface obtain their magnetic moment in this state
due to spin-polarization by the two upper Mn layers.

We also observe a strong coupling of magnetic and geometric structure, with a tight binding for antiferromagnetic coupling between the two upper layers and a larger interlayer distance for a ferromagnetic alignment.
Note, that the strong inward relaxation of the top Mn layer in the
$\uparrow\downarrow\downarrow$ state is consistent with the small
apparent height difference between the Mn DL and 3L (cf.~Fig.~\ref{fig:overview}).

Our result for the Mn triple layer is consistent with a DFT study for the Mn 
double layer on W(001) 
which also indicated a preferred ferromagnetic coupling between adjacent layers \cite{meyer2020}. 
Due to the $c(2\times 2)$  antiferromagnetic state in the surface Mn layer of the double layer, 
the magnetic moment of the interfacial Mn layer is quenched in Mn-DL/W(001), 
indicating that it is also only induced by the adjacent Mn layer \cite{meyer2020}. 
The antiferromagnetic coupling between magnetic moments of the upper Mn layers 
is also consistent with the DFT calculations of Dennler \textit{et al.}~\cite{dennler2005}. 
The same type of coupling occurs in $\delta$-Mn due to the weakened influence of the W substrate.

\begin{figure}
    \centering
    \includegraphics[width=0.485\textwidth]{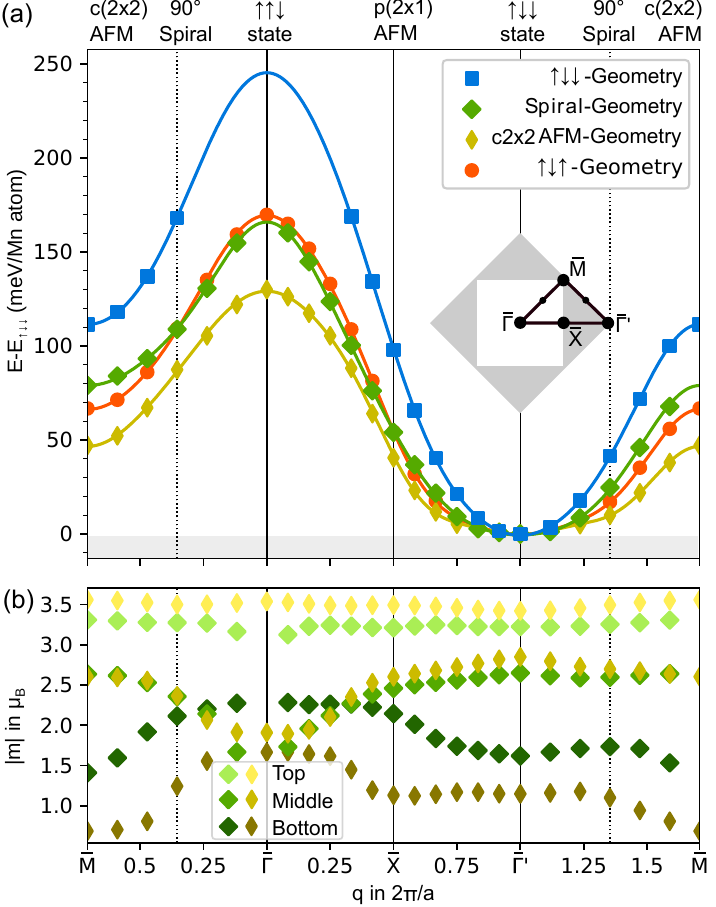}
    \caption{(a) Energy dispersion $E(\mathbf{q})$
    of spin spirals propagating in the Mn triple layer on W(001). Data points represent DFT total energies,
    while connecting lines are a fit to the atomistic spin model. Calculations were carried out for the relaxed geometries (cf.~Tab.~\ref{tab:relaxation})
    of four different magnetic states: the $\uparrow\downarrow\downarrow$ state (blue squares), the 
    $90^\circ$ spin spiral state along $\overline{\Gamma'\text{M}}$ 
    (green diamonds), the c($2 \times 2$) AFM state (yellow diamonds), and
    the $\uparrow\downarrow\uparrow$ (LAFM) state (orange circles).
    The spin spiral vectors \textbf{q} are chosen along the high-symmetry directions of the first (white square) and the second (gray diamond) Brillouin zone (see inset). (b) Magnetic moments for each Mn layer of a spin spiral with the ground state geometry of the $90^\circ$ spin spiral
    along $\overline{\Gamma'\text{M}}$ (green) and the $c(2\times 2)$ antiferromagnet (yellow).}
    \label{fig:dispersion}
\end{figure}

{\bf Spin spiral calculations.}
We have shown that the magnetic contrasts observed in the SP-STM images can 
be explained by
a spin spiral state with an angle of about $87.5^\circ$ (Fig.~\ref{fig:STMsimulation}).
Therefore, we have performed DFT total energy calculations for spin spiral states. 
Spin spirals are spatially rotating magnetic structures characterized 
by a spiral vector $\mathbf{q}$ in reciprocal space (see Fig.~\ref{fig:magneticStates}(f) and (h)). 
The magnetic moment $\mathbf{m}_i(\mathbf{q})$
at site $i$ 
at the position $\mathbf{R}_i$ 
is given by 
\begin{equation}
    \mathbf{m}_i(\mathbf{q})=m_{L(i)} 
    \left[
    \cos(\mathbf{R}_i\cdot\mathbf{q})\cdot\hat{\mathbf{z}}
    +
    \sin(\mathbf{R}_i\cdot\mathbf{q})\cdot\hat{\mathbf{q}}
    \right],
    \label{eq:spinspiral}
\end{equation}
where $m_{L(i)}$ is the magnitude of the magnetic moment 
in layer $L(i)\in\{\text{T},\text{M},\text{B}\}$ (see Tab. \ref{tab:relaxation}).
Eq.~(\ref{eq:spinspiral})
describes a cycloidal spin spiral that rotates in the $zq$-plane.

High-symmetry points of the two-dimensional Brillouin zone 
(2D BZ, inset of Fig.~\ref{fig:dispersion}(a)) correspond 
to collinear magnetic states. 
At the $\overline{\Gamma}$-point, we find the 
$\uparrow \uparrow \downarrow$ state (Fig.~\ref{fig:magneticStates}(b)),
at the $\overline{\Gamma}'$-point, the $\uparrow \downarrow \downarrow$
state (Fig.~\ref{fig:magneticStates}(c)), at the $\overline{\rm M}$-point
the c$(2 \times 2)$ AFM state (Fig.~\ref{fig:magneticStates}(e))
and at the $\overline{\rm X}$-point the p$(2 \times 1)$ AFM state.
For a spin spiral vector $\mathbf{q}$ half-way along the high-symmetry direction
$\overline{\Gamma'\rm M}$ 
one obtains a spin spiral with angles of $90^\circ$ between
adjacent Mn moments within a layer (Fig.~\ref{fig:magneticStates}(h))
which could explain the SP-STM measurements. 
Note, that the orientation of the plane in which the spins rotate is
not fixed as long as spin-orbit coupling is neglected. 

Starting from the energetically preferred $\uparrow\downarrow\downarrow$ state
at the $\overline{\Gamma}'$ point of the 2D BZ, $\textbf{q}$ was varied for a spiral propagating in all Mn layers which results in the
energy dispersion $E(\textbf{q})$ shown in Fig.~\ref{fig:dispersion}(a). 
In order to check the influence of the structure of the Mn triple layer, 
spin spiral calculations were performed for four different geometries 
corresponding to the relaxed structure of
the $\uparrow\downarrow\downarrow$-state, the $90^\circ$ spin spiral along $\overline{\Gamma' \rm M}$, a c($2 \times 2$) antiferromagnetic state in all magnetic layers and the $\uparrow\downarrow\uparrow$-state (cf.~Table \ref{tab:relaxation}). Interlayer distances for all geometries were taken from the {\tt FLEUR} calculations, except for the $90^\circ$ spin spiral geometry, which was taken from the {\tt VASP} calculation.

The energy minimum of $E(\textbf{q})$ lies for all geometries at the $\overline{\Gamma}$'-point, 
i.e.~at the $\uparrow\downarrow\downarrow$ state (Fig.~\ref{fig:dispersion}(a)). 
The spin spiral energy increases as $\textbf{q}$ is varied along the
$\overline{\Gamma' \text{X}}$ and $\overline{\Gamma' \text{M}}$
direction of the two-dimensional Brillouin zone and reaches its maximum at the $\overline{\Gamma}$-point of the first Brillouin zone, i.e.~at the 
$\uparrow\uparrow\downarrow$ state. However, the curvature of $E(\textbf{q})$
in the vicinity of $\overline{\Gamma}$' depends significantly on the
structural relaxation, i.e.~the Mn interlayer distances. 
The 90° flat spin spiral at $\overline{\Gamma' \rm M}/2$
which can explain the SP-STM images
(Fig.~\ref{fig:STMsimulation}) is not favored in our DFT calculations 
neglecting spin-orbit coupling
over the $\uparrow\downarrow\downarrow$ state 
in any of the investigated geometries. In its own ground state geometry 
its energy is 25 meV/Mn-atom higher than the $\uparrow\downarrow\downarrow$ state (Fig.~\ref{fig:dispersion}(a)). For the structure of the c$(2 \times 2)$ AFM state, 
the energy of the $90^\circ$ spin spiral state is only by 10 meV/Mn atom above
the $\uparrow\downarrow\downarrow$ state.

\begin{table*}[htbp]
    \centering
    \begin{tabular}{cccccccccc}
         \hline \hline
           geometry
            & $J_1^\text{eff}$ 
            & $J_2^\text{eff}$ 
            & $J_3^\text{eff}$ 
            & $J_4^\text{eff}$ 
            & $J_5^\text{eff}$ 
            & $J_6^\text{eff}$ 
            & $J_7^\text{eff}$ 
            & $J_8^\text{eff}$ 
            & $J_9^\text{eff}$ \\\hline
            $\uparrow\downarrow\downarrow$
            & $-$7.85 & $-$0.40 & $-$0.58 &  0.09 & $-$0.09 &    0.04 &    0.03 & $-$0.03 &    0.02 \\
            $\uparrow\downarrow\uparrow$
            & $-$5.50 & $-$0.65 & $-$0.65 &  0.17 & $-$0.07 & $-$0.02 &    0.04 & $-$0.06 &    0.02 \\
            $\overline{\Gamma'\text{M}}$ 
            & $-$5.13 & $-$0.05 & $-$0.81 &  0.05 & $-$0.05 & $-$0.14 & $-$0.04 & $-$0.01 & $-$0.01 \\
            AFM
            & $-$4.39 & $-$0.66 & $-$0.44 &  0.26 & $-$0.01 & $-$0.03 &    0.05 & $-$0.08 &    0.02 \\
         \hline \hline
    \end{tabular}
    \caption{Effective exchange constants, $J_i^\text{eff}$
    for $i$-th neighbors obtained by fitting the energy dispersions of 
    the Mn-TL/W(001) (Fig.~\ref{fig:dispersion}(a)) for different ground state 
    geometries to an effective single magnetic layer spin model. 
     In the table the c$(2 \times 2)$ AFM state has been
     abbreviated by AFM
     and the $90^\circ$ spin spiral along 
     $\overline{\Gamma'\text{M}}$ by $\overline{\Gamma'\text{M}}$.
     The exchange constants are given in meV.}
    \label{tab:effectiveparameters}
\end{table*}

The energy dispersion $E(\textbf{q})$ calculated via DFT for spin spirals 
neglecting spin-orbit coupling (Fig.~\ref{fig:dispersion}(a))
can be used to parameterize an effective atomistic 
spin model including only Heisenberg exchange interactions.
For simplicity, the triple layer is modeled as a single magnetic layer with one type of magnetic atoms. The spins in the model are placed in a plane at the $xy$-coordinates of the Mn atoms in the trilayer. Thereby, the top and bottom
layer Mn atoms are mapped to a single site, while the middle layer represents
another site of the effective square spin model with a nearest-neighbor 
distance of $a/\sqrt{2}$.
Since there is only one species of atoms in the effective spin model, interactions of atoms that have the same distance within the $xy$-plane, i.e.~Mn atoms from
the top and bottom layer, will be added together. The predictions of the effective model for each energy dispersion are
displayed as the continuous colored lines in Fig.~\ref{fig:dispersion}(a). 
The parameters of the effective model are listed in 
Tab.~\ref{tab:effectiveparameters}.

\begin{figure}
    \centering
    \includegraphics[width=0.485\textwidth]{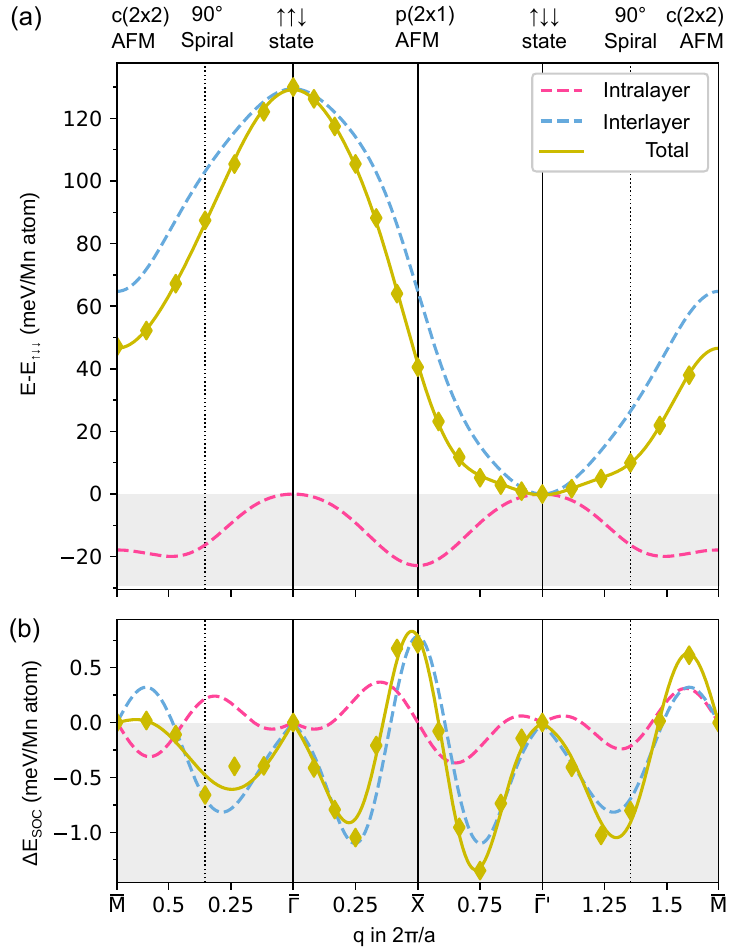}
    \caption{
    Energy dispersion $E(\mathbf{q})$ of flat cycloidal spin spirals 
    in the Mn-TL/W(001)
    for the ground state geometry of the $c(2\times2)$ AFM state.
    (a) Spin spiral energy without spin-orbit coupling (SOC).
    DFT total energies (points) are fitted with a model of bilinear exchange (yellow line).
    Contributions of effective intralayer and interlayer exchange in the model
    are displayed in pink and blue, respectively.
    (b) SOC contribution to the energy dispersion of 
    a cycloidal, counter-clockwise rotating spin spiral,
    in first order perturbation theory. The DFT data is fitted with 
    the DMI of the effective spin model, 
    which can also be separated into intralayer and interlayer contributions.
    Positive energy values imply the preference of a clockwise spiral by the same amount.
    Note, that the interlayer interactions between the top and the 
    bottom Mn layer 
    are mapped onto effective intralayer interaction in (a) and (b).
    }
    \label{fig:dmi}
\end{figure}

The exchange interaction between nearest neighbors is the dominating interaction in the spin model of the film ranging from a maximum antiferromagnetic
strength of $J_1^\text{eff}=-7.85$ meV for the $\uparrow\downarrow\downarrow$-ground state geometry to a minimal strength of
$J_1^\text{eff}=-4.39$ meV for the c$(2 \times 2)$ AFM ground state geometry. 
This interaction is the sum of the interactions between nearest neighbors 
in the top with middle and middle with bottom layers in the trilayer,
making it an interlayer interaction. 
This result is consistent with the 
above investigation of interlayer coupling based on
collinear magnetic states (cf.~Table \ref{tab:relaxation}), in which the antiferromagnetic coupling between top and middle layer dominates the ferromagnetic coupling 
between the bottom two layers.
The most preferable magnetic state for a negative $J_1^\text{eff}$ is a $c(2 \times2)$ antiferromagnetic state in the effective
spin model, which translates to a triple layer with 
ferromagnetically aligned layers that have an
antiferromagnetic coupling between the layers
consistent with the $\uparrow\downarrow\downarrow$ state.
The next-nearest exchange interaction, $J_2^\text{eff}$, 
also prefers an antiferromagnetic coupling ($J_2^\text{eff}<0$).
It represents the nearest-neighbor intralayer interaction, 
and the next-nearest neighbor interaction between the top and bottom Mn layers.
$J_2^\text{eff}$ competes with $J_1^\text{eff}$, 
since the ferromagnetic state in each layer that is stabilized by $J_1^\text{eff}$ 
is energetically the least optimal state for a negative $J_2^\text{eff}$. 
For a larger ratio $J_2^\text{eff}/J_1^\text{eff}$, 
a spin spiral ground state is expected for the atomistic spin model. 
The same holds for $J_3^\text{eff}$, which is similar to $J_2^\text{eff}$ in sign and strength.
Exchange constants beyond 
$J_3^\text{eff}$
are quite small 
in comparison to $J_1^\text{eff}$ 
and have negligible influence of the magnetic ground state of the spin model, 
leading to checkerboard antiferromagnetic pattern that represents the 
$\uparrow\downarrow\downarrow$ in the real system.
For the c$(2\times 2)$ AFM ground state geometry, 
the competing contributions of interlayer and intralayer exchange 
to the energy dispersion $E(\mathbf{q})$
can be seen in Fig.~\ref{fig:dmi}(a).
In Fig.~\ref{fig:dmi}(a) it is apparent that the
two types of exchange interactions compete as the intralayer exchange 
prefers a spin spiral state, while the interlayer exchange favors the collinear 
$\uparrow\downarrow\downarrow$ state.

The 90° spin spiral can be used to create another state that is 
also consistent with the $c(4\times2)$ magnetic unit cell suggested by the
SP-STM measurements:
a superposition of two 90° spin spirals 
with opposite rotational sense leading to the $uudd$ state \cite{Hardrat2009} (see Fig.~\ref{fig:magneticStates}(g)). 
This state has recently been observed in hexagonal magnetic monolayers on 
surfaces \cite{Kroenlein2018,Romming2018}.
On a bcc (001) surface, it exhibits a $c(4\times2)$ magnetic unit cell.
It is degenerate in energy with the 90° spin spiral within the Heisenberg model
of pair-wise exchange. Total energy differences obtained by DFT calculations
must stem from 
higher-order interactions \cite{Hardrat2009,Kroenlein2018,Hoffmann2020}. 
We calculated the total energies of an $uudd$ state and a 90° spin spiral 
with
the 
{\tt VASP} code in a $c(4\times2)$ supercell using
the relaxed geometry of the $\uparrow\downarrow\downarrow$-state.
The obtained energies of 37 meV/Mn-atom for the 90° spiral
with respect to the $\uparrow\downarrow\downarrow$ state, which is consistent 
with the {\tt FLEUR} result, and 60 meV/Mn-atom for the $uudd$ state 
with respect to the $\uparrow\downarrow\downarrow$ state show that 
higher-order interactions favor the 90° spin spiral over the $uudd$ state. 
Note, that the $uudd$ state differs from the $87.5^\circ$ spin spiral state
in that its SP-STM image only shows a zigzag pattern and cannot explain the 
observed stripe contrast (cf.~Fig.~\ref{fig:STMsimulation}). 

{\bf Dzyaloshinskii-Moriya interaction.}
Since the DMI is known to favor spin spirals over the ferromagnetic state,
we investigated if the inclusion of SOC could lead to a 
non-collinear ground state in the Mn trilayer.
The contributions of SOC to the total energies of spin spirals
were calculated using {\tt FLEUR} within first order perturbation theory, 
starting from self-consistent cycloidal spin spiral states in the $c(2\times 2)$ AFM 
ground state geometry (see Fig.~\ref{fig:dmi}(b)).
Negative energy values decrease the energy of counter-clockwise rotating spirals 
and positive values decrease the energy of clockwise rotating ones.

Energy contributions at the $\overline{\rm X}$-point indicate the presence of 
interlayer DMI, since each layer exhibits a row-wise antiferromagnetic state at 
this high-symmetry point, i.e.~a collinear state without intralayer DMI contributions.
By mapping the energies to an effective monolayer system, as for the exchange interaction,
it can be revealed that the interlayer DMI is the dominating term. Both interlayer
and intralayer DMI favor spin spiral states with an energy minimum along the 
$\overline{\Gamma' {\rm X}}$ direction. Along the $\overline{\Gamma' {\rm M}}$ direction the energy minimum is close to the $90^\circ$ spin spiral.
However, SOC contributions are two orders of magnitude smaller than the total energy 
of flat spin spirals without SOC.

{\bf Magnetic anisotropy.}
We obtained only a very small magnetocrystalline anisotropy 
energy of 0.01~meV/Mn atom for the $\uparrow\downarrow\downarrow$ state
of the Mn triple layer on W(001)
favoring an in-plane orientation of the magnetization.
We attribute the small SOC effect to the fact that
the bottom Mn layer at the interface to the W substrate exhibits only
a small magnetic moment which is induced by the adjacent central Mn layer. Therefore, the
Mn layers with significant intrinsic magnetic moments
are at a relatively large distance from the heavy W substrate needed to create
large SOC effects. 

To compute the influence of the magnetic dipole-dipole 
interactions we numerically summed over the magnetic fields of 
atomic magnetic moments with a distance of up to 1000 
in-plane lattice constants taking the geometrical structure
and the magnetic moments of the three different Mn layers 
obtained from DFT for the $\uparrow\downarrow\downarrow$ state 
into account. From these calculations we found that the magnetostatic
dipolar interaction favors an in-plane orientation of the magnetic moments
in the $\uparrow\downarrow\downarrow$ state by 0.036~meV/Mn atom with
respect to an out-of-plane orientation. This small value can be 
understood from the net magnetic moment of the Mn triple layer 
which amounts to only about 0.75~$\mu_{\rm B}$ per unit cell
(cf.~Table \ref{tab:relaxation}).
The inclusion of SOC does not change the ground state of the system. 
DMI alone would prefer a spin spiral with a period of about 200 lattice constants 
by 4~$\mu$eV close to the $\uparrow\downarrow\downarrow$-state.
However, the magnetocrystalline anisotropy raises the energy of all cycloidal spin
spiral states by 6~$\mu$eV, thus stabilizing the collinear ground state.

\begin{figure}
    \centering
    \includegraphics[width=0.485\textwidth]{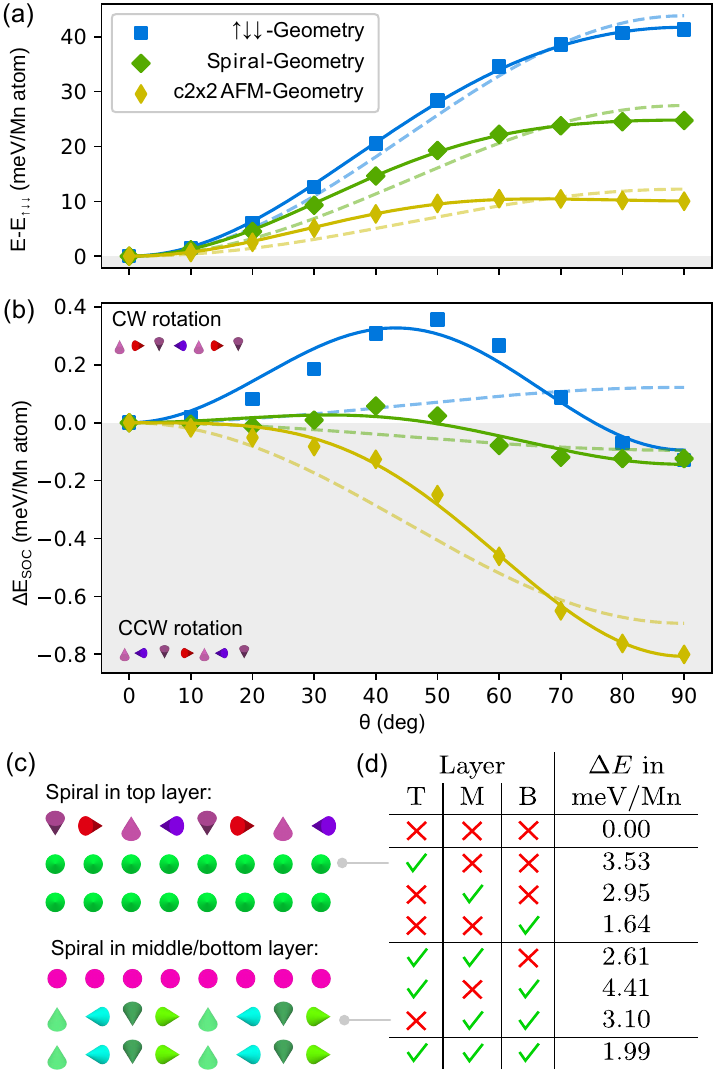}
    \caption{
    (a) Total DFT energies of conical 90° spin spirals in the Mn-TL/W(001)
    along the $\overline{\Gamma'\text{M}}$
    direction as a function of the opening angle $\theta$ neglecting SOC. 
    The dashed line represents a fit with only Heisenberg exchange,
    while the solid line includes also the biquadratic exchange 
    (see Eq.~(\ref{eq:conical})). 
    All calculations were performed in three different ground-state geometries
    as indicated.
    (b) SOC contributions to the energy of conical spin spirals
    obtained in first-order perturbation theory.
    The dashed line represents a fit with only DMI, i.e.~first term of Eq.~(\ref{eq:conicalSOC}),
    while the solid line includes also the chiral biquadratic pair interaction, i.e.~both
    terms of Eq.~(\ref{eq:conicalSOC}). 
    (c) Sketch of a conical spin spirals in selected Mn layers. 
    Ferromagnetic layers point towards the viewer 
    while spiral parts rotate in a perpendicular direction.
    (d) Total DFT energies without SOC of conical spin spirals propagating only in selected Mn layers, with an opening angle of $\theta=10^\circ$. The first three columns of the table indicate in which of the layers the spirals propagate (top, middle and bottom layer).
    }
    \label{fig:conical}
\end{figure}

{\bf Conical spin spiral states.}
We can obtain spin spiral states of lower energy than the flat $90^\circ$ spin 
spiral by
superimposing it with the $\uparrow\downarrow\downarrow$-state at each lattice site $i$,
thereby constructing a conical spin spiral states with magnetic moments
\begin{equation}
\label{eq:conicalconstruct}
    \mathbf{m}_i(\mathbf{q},\theta) = 
    \sin(\theta)\cdot\mathbf{m}_i(\mathbf{q})
    +
    \cos(\theta)\cdot m_{L(i)}(\hat{\mathbf{q}}\times \hat{\mathbf{z}})
    ,
\end{equation}
which is characterized by
the opening angle $\theta$ that its moments and the rotational axis $\hat{\mathbf{q}}\times \hat{\mathbf{z}}$
enclose (Fig.~\ref{fig:magneticStates}(i)).
The total energy of a conical spin spiral can be described within 
the extended Heisenberg model by
\begin{equation}
    E = 
    - \sum_{ij} J_{ij}   \left(\textbf{m}_i\cdot \textbf{m}_j \right)
    - \sum_{ij} B_{ij} \left(\textbf{m}_i\cdot \textbf{m}_j \right)^2,
\end{equation}
in which a biquadratic exchange term has been included
as a higher-order interaction to the bilinear exchange.
By computing the energy of the conical spin spiral from 
Eq.~(\ref{eq:conicalconstruct}), the energy 
\begin{equation}
\label{eq:conical}
    E_\mathbf{q}(\theta) - 
    E_\text{$\uparrow\downarrow\downarrow$} = 
     J^\text{eff}_\mathbf{q} \sin^2(\theta) 
    + B^\text{eff}_\mathbf{q} \sin^4(\theta)
\end{equation}
of a spiral with the vector $\mathbf{q}$ as a function of the opening angle can be derived.
For small $\theta$ the energy comes close to the energy $E_\text{$\uparrow\downarrow\downarrow$}$ of the $\uparrow\downarrow\downarrow$ state.
The first and second term arise from bilinear and biquadratic exchange, respectively.
The addition of higher-order exchange interactions such as the biquadratic term
can lead to an energy minimum
for a conical spin spiral as reported for a Mn DL on W(110) \cite{Yoshida2012}.

For the 90° spin spiral near the $\overline{\Gamma}^\prime$-point, 
$E_\mathbf{q}(\theta)$ was calculated via DFT for three different ground state geometries
(Fig.~\ref{fig:conical}(a)).
The obtained total DFT energies closely follow a $\sin^2(\theta)$ dependence, 
indicating large contributions from bilinear exchange, i.e.~the first term
on the right hand side 
of Eq.~(\ref{eq:conical}). 
Higher-order exchange contributions significantly improve
the fit of the DFT data (Fig.~\ref{fig:conical}(a)).
However, they are comparatively small and tend to favor flat spin spirals.

To investigate the influence of SOC, we have calculated the energy contribution due to SOC on the energy 
of conical spin spirals in the same computational setup using first-order perturbation theory via the
{\tt FLEUR} code.
To model the effect of SOC on the energy within the atomistic spin model we used 
\begin{equation}
\begin{split}
    \Delta E_\text{SOC}=& 
    - \sum_{ij} \mathbf{D}_{ij}   \left(\textbf{m}_i\times \textbf{m}_j \right)\\
    &- \sum_{ij} \mathbf{C}_{ij} \left(\textbf{m}_i\times \textbf{m}_j \right)
    \cdot \left(\textbf{m}_i\cdot \textbf{m}_j \right),
\end{split}
\end{equation}
which contains the DMI as well as the chiral biquadratic pair-interaction 
that was recently proposed in Ref.~\cite{Brinker19}.
The two vectors $\mathbf{D}_{ij}$ and $\mathbf{C}_{ij}$  are assumed to align perpendicular to
the plane of rotation of the spiral.
For the conical spin spiral, Eq.~(\ref{eq:conicalconstruct}), the SOC energy
\begin{equation}
\label{eq:conicalSOC}
    \Delta E_{\text{SOC},\mathbf{q}}(\theta) = 
    D^\text{eff}_\mathbf{q} \sin^2(\theta) 
    + C^\text{eff}_\mathbf{q} \sin^4(\theta)
\end{equation}
has the same dependence on $\theta$ as the exchange energy.
The second term arises because of the chiral biquadratic pair interaction.

The energy contribution $\Delta E_{\text{SOC},\mathbf{q}}(\theta)$ calculated via DFT
(Fig.~\ref{fig:conical}(b)) depends strongly on
the geometry of the film system, i.e.~the Mn interlayer distances.
For the $c(2\times 2)$ AFM ground state geometry, 
the highest SOC contribution is computed for the flat spin spiral 
($\theta=90^\circ$) with about 0.8~meV/Mn atom, preferring an 
counter-clockwise rotation. The fit of the DFT data to Eq.~(\ref{eq:conicalSOC})
shows that the
chiral biquadratic term plays a non-negligible role
(Fig.~\ref{fig:conical}(b)). For the other two considered
ground state geometries, 
the SOC energy for the flat spin spiral ($\theta=90^\circ$) 
decreases drastically to only 0.1~meV/Mn atom.
However, for an opening angle $\theta$ of about $40$ to $50^\circ$ we
observe a change of the preferred rotational sense
which cannot be explained by the DMI. This unexpected dependence
of the SOC energy for conical spin spirals as a function of
the opening angle stems from higher-order interactions due to SOC 
such as the chiral biquadratic term.
Especially, for the $\uparrow\downarrow\downarrow$ ground state
geometry there is a significant energy gain of about 0.4 meV/Mn atom
for a clockwise rotating conical spin spiral with an opening angle
of $50^\circ$. Nevertheless, the SOC contributions are at least one 
order of magnitude smaller, even in the $c(2\times 2)$ AFM ground 
state geometry. Therefore, within our DFT calculations
SOC does not lead to a 90° conical spin spiral 
that would explain the experimental results.
    
We have also investigated whether the energy of the conical 90° spin
spiral with an opening angle of $\theta = 10^\circ$
can be lowered by propagating it only in one or two of the Mn layers, while the other layers are in a ferromagnetic state (sketch in Fig.~\ref{fig:conical}(c)).
The results of these calculations are collected in Fig.~\ref{fig:conical}(d). 
All total energies are given with respect to the $\uparrow\downarrow\downarrow$ state
and have been obtained neglecting SOC.
While the creation of a spin spiral in one magnetic layer requires energy, 
less energy is needed for the creation of a second spiral. If both spin
spirals propagate in the two top layers, it even lowers the energy with
respect to the case of only one spin spiral.
This can be explained by the strong antiferromagnetic coupling between the top two layers. The state with spin spirals in all three Mn layers possesses the lowest energy. Such
a conical 90° spin spiral with an opening angle of $\theta=10^\circ$ is only about 2~meV/Mn atom higher than the $\uparrow\downarrow\downarrow$ state. Note,
that the energy differences of conical spin spirals with respect to the
$\uparrow\downarrow\downarrow$ state are expected to be even lower if 
the calculations were performed for a different structural relaxation
similar to the result obtained for flat spin spirals 
(cf.~Fig.~\ref{fig:conical}(a)).

\section{Conclusion}

In our SP-STM experiments, we found the well-known mesh-like $ \surd 2 \times \surd 2$ structure 
on the Mn DL on W(001), which was first described in Ref.~\citep{meyer2020}. 
This result is in perfect agreement with the c($2 \times 2$) AFM order in the surface Mn layer 
and quenched magnetic Mn moments at the Mn/W interface reported by Meyer {\em et al.} \cite{meyer2020}. 
Our SP-STM data of pseudomorphic 3L Mn/W(001) films show periodic zigzag patterned stripes 
consistent with a magnetic $2\surd 2 \times \surd 2$ unit cell.  
The magnetic origin of this pattern was verified by experiments, where the tip magnetization 
was reversed at close tip--sample distance via the exchange interaction,  
resulting in a characteristic contrast inversion for both DL and 3L Mn, see Fig.~\ref{fig:contrastreversal}.  
A limitation of these spin-polarized measurements lies in the uncertainty 
of the magnetization direction of the spin-polarized Mn/W tip,   
namely whether it is entirely out-of-plane magnetized or also includes a significant in-plane component.

SP-STM simulations revealed that a spin spiral state with an angle close to $90^\circ$ can explain the observed magnetic contrast. A flat spin spiral as well as a conical spin spiral with a finite opening angle is consistent
with the SP-STM experiments since the magnetization direction of the tip could not be fully determined.

DFT calculations have been performed using the {\tt FLEUR} as well as the {\tt VASP} code
to shed light on the magnetic ground state and its origin.
Our calculations show that the geometric structure of the Mn triple layer and its magnetic state 
are closely linked. In particular, the interlayer distances between the Mn layers and to the W substrate change considerably upon changing the magnetic state in the triple-layer.

We found a novel collinear magnetic state for the Mn triple-layer on W(001) which is energetically much more favorable in DFT than the layered antiferromagnetic state previously proposed by Dennler {\it et al.} \cite{dennler2005}.
In this so-called $\uparrow \downarrow \downarrow$ state the magnetic moments of the Mn surface and central
layers are oppositely aligned with respect to each other. The magnetic moments of the Mn atoms in the
interface layer to the W substrate are aligned parallel to those of the central layer and only induced by 
the central Mn layer.

Spin spiral calculations performed starting from the $\uparrow \downarrow \downarrow$ state show
a total DFT energy rise upon canting the spins, i.e.~deviating from the collinear state. 
Therefore, a flat $90^\circ$ spin spiral state is
unfavorable with respect to the  
$\uparrow \downarrow \downarrow$ state. However, the energy difference depends significantly
on the interlayer distances in the Mn triple-layer. 
We find that the interlayer exchange prefers the collinear state, while
intralayer exchange interactions favor a spin spiral state. The DMI prefers a $90^\circ$
cycloidal spin spiral state, however, its energy contribution is much smaller than that of
the exchange. The magnetocrystalline anisotropy energy is also small in the 
$\uparrow\downarrow\downarrow$ state and favors an in-plane magnetization. We attribute
the small size of the SOC effects to the small Mn moments at the W interface in 
the $\uparrow\downarrow\downarrow$ state.

DFT calculations for conical spin spiral states show that the energy still rises upon deviating 
from the collinear $\uparrow \downarrow \downarrow$ state, however, the energy difference is much reduced. Based on these results, we can exclude a significant effect of higher-order exchange interactions in the Mn triple-layer which could be responsible for a conical spin spiral as found 
for a Mn DL on W(110) \cite{Yoshida2012}. 
Surprisingly, we found that the energy contribution
due to SOC for conical spin spirals deviates qualitatively from that expected by the DMI.
Therefore, higher-order interactions due to SOC such as the chiral biquadratic pair interaction
need to be taken into account. Such interactions might explain the occurrence of a conical spin
spiral state in the Mn triple-layer since
the energy difference between a $90^\circ$ conical spin spiral 
state with a small opening angle of $10^\circ$ -- which can explain the
observed SP-STM images -- is only slightly higher in energy than the 
$\uparrow \downarrow \downarrow$ state. 
Different opening angles of the conical spin spiral
state in the three Mn layers not considered in our DFT calculations
might also further lower its total energy.

\begin{acknowledgments}
We would like to thank S.\,Meyer for fruitful discussions and S.\,Haldar for
technical support with the {\tt VASP} calculations.
We acknowledge support by DFG through W{\"u}rzburg-Dresden Cluster of Excellence (ct.qmat). T.~D.~and S.~H.~gratefully acknowledge financial support from the Deutsche Forschungsgemeinschaft (DFG, German Research Foundation) via the
SPP2137 "Skyrmionics" (project no.~462602351) and computing time provided by 
the North-German Supercomputing Alliance (HLRN).

\end{acknowledgments}

\end{document}